# Quantum Gravity :
# Has Spacetime Quantum Properties ? [1]

**Reiner Hedrich**[2]

Institut für Philosophie
Fakultät Humanwissenschaften und Theologie
Technische Universität Dortmund
Emil-Figge-Strasse 50
44227 Dortmund
Germany

Zentrum für Philosophie und Grundlagen der Wissenschaft
Justus-Liebig-Universität Giessen
Otto-Behaghel-Strasse 10 C II
35394 Giessen
Germany

**Abstract**

The conceptual incompatibility between General Relativity and Quantum Mechanics is generally seen as a sufficient motivation for the development of a theory of Quantum Gravity. If – so a typical argumentation – Quantum Mechanics gives a universally valid basis for the description of the dynamical behavior of all natural systems, then the gravitational field should have quantum properties, like all other fundamental interaction fields. And, if General Relativity can be seen as an adequate description of the classical aspects of gravity and spacetime – and their mutual relation –, this leads, together with the rather convincing arguments against semi-classical theories of gravity, to a strategy which takes a quantization of General Relativity as the natural avenue to a theory of Quantum Gravity. And, because in General Relativity the gravitational field is represented by the spacetime metric, a quantization of the gravitational field would in some sense correspond to a quantization of geometry. Spacetime would have quantum properties.

---

[1] Research for this paper was generously supported by the *Fritz-Thyssen-Stiftung für Wissenschaftsförderung* under the project *Raumzeitkonzeptionen in der Quantengravitation*. Thanks also to Brigitte Falkenburg!

[2] Email: Reiner.Hedrich@phil.uni-giessen.de & hedrich@fb14.uni-dortmund.de

But, this direct quantization strategy to Quantum Gravity will only be successful, if gravity is indeed a fundamental interaction. Only if it is a fundamental interaction, the given argumentation is valid, and the gravitational field, as well as spacetime, should have quantum properties. – What, if gravity is instead an intrinsically classical phenomenon? Then, if Quantum Mechanics is nevertheless fundamentally valid, gravity can not be a fundamental interaction; a classical and at the same time fundamental gravity is excluded by the arguments against semi-classical theories of gravity. An intrinsically classical gravity in a quantum world would have to be an emergent, induced or residual, macroscopic effect, caused by a quantum substrate dominated by other interactions, not by gravity. Then, the gravitational field (as well as spacetime) would not have any quantum properties. And then, a quantization of gravity (i.e. of General Relativity) would lead to artifacts without any relation to nature.

The serious problems of all approaches to Quantum Gravity that start from a direct quantization of General Relativity (e.g. non-perturbative canonical quantization approaches like *Loop Quantum Gravity*) or try to capture the quantum properties of gravity in form of a 'graviton' dynamics (e.g. *Covariant Quantization*, *String Theory*) – together with the, meanwhile, rich spectrum of (more or less advanced) theoretical approaches to an emergent gravity and/or spacetime – make this latter option more and more interesting for the development of a theory of Quantum Gravity. The most advanced emergent gravity (and spacetime) scenarios are of an information-theoretical, quantum-computational type. A paradigmatic model for the emergence of gravity and spacetime comes from the *Pregeometric Quantum Causal Histories* approach.

# 1. Introduction: The Mutual Incompatibility of General Relativity and Quantum Mechanics / Quantum Field Theory

The decisive motivation for the development of a theory of Quantum Gravity is generally seen in the mutual conceptual incompatibility between General Relativity on the one hand and Quantum Mechanics and Quantum Field Theory on the other hand. The following three crucial points should elucidate this situation:

(1) General Relativity, today our best theory of gravity as well as of spacetime, treats the gravitational field as a classical dynamical field, represented by the (pseudo-)Riemannian metric of spacetime.[3] But, according to Quantum Mechanics, dynamical fields have quantum properties. So, it seems reasonable to assume the necessity of a (direct or indirect) quantization of the gravitational field. An additional motivation for the quantization of gravity comes from rather conclusive arguments against semi-classical modifications of the Einstein field equations that treat gravity classically and everything else quantum mechanically.[4]

---

[3] All other fields as well as matter are also treated classically by General Relativity.
[4] Cf. Kiefer (1994, 2004, 2005), Peres / Terno (2001), Terno (2006), Callender / Huggett (2001a, 2001b).

*"The right-hand side of the field equations [of general relativity] describes matter sources, the behaviour of which is governed by quantum theory. The left-hand side of the field equations describes gravitation as a classical field. If the right-hand side represents quantized matter then the field equations as they stand are inconsistent."* (Riggs (1996) 2)

(2) In General Relativity, the gravitational field is represented by the metric of spacetime. Therefore, a quantization of the gravitational field would correspond to a quantization of the metric of spacetime. The quantum dynamics of the gravitational field would correspond to a dynamical quantum spacetime. But, Quantum Field Theories presuppose a fixed, non-dynamical background spacetime for the description of the dynamics of quantum fields. They are conceptually inadequate for a description of a dynamical quantum geometry. So, a quantum theory of the gravitational field can scarcely be a Quantum Field Theory, at least not one in the usual sense. – But it is not only the dynamical character of general relativistic spacetime which makes traditional quantum theoretical approaches problematic. The active diffeomorphism invariance[5] of General Relativity is fundamentally incompatible with any fixed background spacetime.[6]

(3) In General Relativity, time is a component of dynamical spacetime. It is dynamically involved in the interaction between matter/energy and the spacetime metric. It can be defined only locally and internally; there is no global time.[7] Quantum Mechanics, on the other hand, treats time as a global background parameter, not even as a physical observable represented by a quantum operator.

## 2. The Quantization of Gravity and the Quantum Nature of Spacetime – The Naive View

If we assume that, fundamentally, all natural systems are quantum systems, and that gravity is a universal interaction with influence on all natural systems, then the conceptual incompatibility of Quantum Mechanics and General Relativity leads to a severe problem for our description of nature. Under these conditions, it is natural to assume that at least one of our actually most fundamental, well-established, but mutually incompatible physical theories is only an approximation to a more fundamental physical description of nature. – But, what would be the most promising way to a construction of such a more fundamental physical theory?

Taken into account the successful experiences with the implementation of all other fundamental interactions into a quantum mechanical description, leading to the (at least empirically) successful Standard Model of Quantum Field Theories, the most natural way to get to a theory of Quantum Gravity seems to be a more or less direct quantization of the gravitational field. If Quantum Mechanics gives a fundamental and universally valid basis for the description of the dynamical behavior of all natural systems, the gravitational field should have quantum properties like all other fundamental interaction fields.[8] These quantum properties of the gravitational field should be subject of the searched-for theory of Quantum Gravity.

The additional and completely rational assumption that General Relativity can be seen as an adequate description of the classical aspects of gravity and spacetime – and their mutual relation –, leads then, together with the arguments against semi-classical theories of gravity, to a strategy which consists basically in a quantization of General Relativity as a natural avenue to a theory of Quantum Gravity.

---

[5] Cf. Earman (2006, 2006a).
[6] Cf. Earman (1986, 1989, 2002, 2006, 2006a), Earman / Norton (1987), Norton (1988, 1993, 2004).
[7] The active diffeomorphism invariance of General Relativity leads not at least to the *Problem of Time*. See Ch. 5 as well as Belot / Earman (1999), Earman (2002), Pons / Salisbury (2005), Rickles (2005), Rovelli (2001, 2006), Isham (1993), Unruh / Wald (1989).
[8] This argument is only valid, if gravity is a fundamental interaction. See Ch. 6 and Ch. 7.

In General Relativity the gravitational field is represented by the metric of spacetime. Gravity is identical to properties of a dynamical geometry. Therefore, a quantization of the gravitational field would correspond to a quantization of the metric of spacetime. The quantum dynamics of the gravitational field would correspond to a dynamical quantum spacetime, a dynamical quantum geometry. A theory of Quantum Gravity should then be or lead to a description of quantum space-time.

> *"[...] general relativity is not just a theory of gravity – in an appropriate sense, it is also a theory of spacetime itself; and hence a theory of quantum gravity must have something to say about the quantum nature of space and time."* (Butterfield / Isham (2001) 34)

But what would we have to expect with regard to the quantum properties of spacetime? – Taken again into account the experiences with Quantum Mechanics, we would probably suspect that the spacetime metric should be the expectation value of a quantum variable. On the quantum level, we would probably expect quantum uncertainties and fluctuations of the spacetime metric as well as superpositions of spacetime metrics. And the experiences with Quantum Field Theories would possibly suggest some exchange boson for gravity: the 'graviton'. Quantum Gravity, one could think, should under these assumptions possibly be a theory describing the dynamics of gravitons exchanged between matter particles.

## 3. Quantum Spacetime – Problems with the Naive View

But this is certainly a much too naive picture, because (i) the apparatus of Quantum Field Theory with its fixed background spacetime is fundamentally incompatible with General Relativity and its active diffeomorphism invariance (background-independence),[9] because (ii) gravitons representing the gravitational field, corresponding to the metric field, describing therefore the quantum dynamics of spacetime, can scarcely be understood as moving within a (miraculously already existing classical) spacetime (required by Quantum Field Theory), and because (iii) gravity can scarcely be seen as an interaction represented simply by field bosons exchanged between matter particles; gravity has an effect on any form of matter and energy.

However, one does not necessarily need to take into account the conceptual problems of the assumption of a graviton dynamics, to see that a simple quantization of General Relativity is probably a very questionable route to a theory of Quantum Gravity. Quantum fluctuations of the spacetime metric, to be expected within the outlined naive picture of a direct quantization of gravity and spacetime, are totally sufficient to lead to serious problems:

> *"[...] once we embark on constructing a quantum theory of gravity, we expect some sort of quantum fluctuations in the metric, and so also in the causal structure. But in that case, how are we to formulate a quantum theory with a fluctuating causal structure?"* (Butterfield / Isham (2001) 64)

Quantum fluctuations of spacetime are fluctuations of the basic causal structure. This leads very probably to insurmountable problems for a direct quantization of the gravitational field in the context of the known quantization procedures.

---

[9] See Ch. 4 for the consequences of a background-dependent quantization of a background-independent theory.

> *"The main idea [...] is that fluctuations in the gravitational field imply fluctuations in the spatiotemporal, and hence causal, structure of the world. But it is hard to see how one can make sense of canonical commutation relations and hence quantize anything in the absence of a stable causal structure."*
> (Callender / Huggett (2001a) 22)

So, fluctuations of the spacetime metric, to be expected within the naive approach to a description of quantum spacetime, are completely sufficient to make clear that one can not get over the mutual incompatibility of General Relativity and Quantum Mechanics by simply applying the usual quantization procedures to the gravitational field. – Possibly fluctuations of the causal structure of spacetime exist. Possibly there does not even exist any basic causal structure in nature. Possibly causal structure is only an approximately valid concept or an emergent phenomenon. But all these imaginable possibilities and speculations about the quantum properties of spacetime will not be elucidated any further within an approach starting from a simple quantization of the gravitational field, inspired by the procedures of Quantum Field Theory. The methodological problems posed by quantum fluctuations of spacetime metric, to be expected within the naive approach to a quantum spacetime, are insurmountable within this naive approach. They lead to implications pointing beyond the context of a simple direct quantization of General Relativity.

At this point, it is reasonably clear that a direct quantization of General Relativity, following the outlined strategy, will not be an adequate route to Quantum Gravity. Such a direct quantization of General Relativity implies the assumption of a fundamental spacetime with additional quantum properties: quantum corrections to a classical spacetime. This assumption leads in turn to problems which make a direct quantization of General Relativity (at least within the traditional strategies of Quantum Mechanics and Quantum Field Theory) conceptually impossible. This suggests that General Relativity and Quantum Mechanics are too different to allow a simple amalgamation. A theory of Quantum Gravity, a theory that gets over their mutual incompatibility, has to be constructed in a different way. – But let us see first (Ch. 4) what happens concretely, when one tries in fact to quantize gravity the naive way, and second (Ch. 5), if there are more sophisticated ways of quantizing General Relativity that lead to the discovery of any loopholes in the foregoing arguments against a direct quantization of gravity.

## 4. *Covariant Quantization* of General Relativity: Graviton Dynamics

The *Covariant Quantization*[10] of General Relativity reflects the problems outlined for the naive picture of a quantum spacetime in a direct and concrete way. *Covariant Quantization* consists in the attempt to construct a Quantum Field Theory of gravity, which means: a Quantum Field Theory of the metric field.

> *"The idea was [...] to do unto the gravitational field as was done to the electromagnetic field: quantize the gravitational field to a get a particle (the* graviton*) that mediates the interaction."* (Rickles / French (2006) 16)

But Quantum Field Theories need a background spacetime with fixed metric for the definition of its operator fields.

> *"However, just as photons require a background metrical structure, so does the graviton."* (Rickles / French (2006) 16)

---

[10] Cf. DeWitt (1967a, 1967b).

Consequently, *Covariant Quantization* uses a standard perturbation-theoretical approach, working with a fixed (usually Minkowski) background metric and a perturbation on this background to be treated quantum mechanically. This leads to a Quantum Field Theory of the fluctuations of the metric. The properties of the corresponding field quanta of gravity are a consequence of symmetry arguments and of the properties of classical gravity: long-range, exclusively attractive. 'Gravitons' are massless and have spin 2. They represent the assumed quantum properties of spacetime, and they behave according to standard Feynman rules on a fixed background spacetime.

> *"Field-theoretic techniques are put at the forefront. The first step in this program is to split the space-time metric $g_{\mu\nu}$ in two parts, $g_{\mu\nu} = \eta_{\mu\nu} + \sqrt{G}\, h_{\mu\nu}$, where $\eta_{\mu\nu}$ is to be a background, kinematical metric, often chosen to be flat, G is Newton's constant, and $h_{\mu\nu}$, the deviation of the physical metric from the chosen background, the dynamical field. The two roles of the metric tensor are now split. The overall attitude is that this sacrifice of the fusion of gravity and geometry is a moderate price to pay for ushering-in the powerful machinery of perturbative quantum field theory. [...] it is only $h_{\mu\nu}$ that is quantized. Quanta of the field propagate on the classical background space-time with metric $\eta_{\mu\nu}$. If the background is in fact chosen to be flat, one can use the Casimir operators of the Poincaré group and show that the quanta have spin two and rest mass zero. [...] Thus, in this program, quantum general relativity was first reduced to a quantum field theory in Minkowski space."* (Ashtekar (2005) 5)

But *Covariant Quantization* with its perturbation expansion of the fluctuations of the spacetime metric turns out to be non-renormalizable. This makes the theory irrelevant as a fundamental description of spacetime.

> *"It is generally agreed that this non-renormalisability renders perturbatively quantised Einstein gravity meaningless as a fundamental theory because an infinite number of parameters would be required to make any physical prediction."* (Nicolai / Peeters / Zamaklar (2005) 3)

The non-renormalizability of the theory is a direct consequence of the self-interaction of the graviton, which is in turn a quantum-field-theoretical expression of the nonlinearity of classical gravity. Gravity couples to mass and, because of the mass-energy equivalence, to every form of energy.[11] Therefore the self-interaction contributions to gravity increase for decreasing distances or increasing energies. So, the contribution of virtual particles with increasing energies dominates the higher orders of the perturbation expansion. This leads to uncontrollable divergences of the expansion and to the non-renormalizability of the perturbative Quantum Field Theory of gravity.

> *"[...] such non-renormalizable theories become pathological at short distances [...] – perhaps not too surprising a result for a theory which attempts in some sense to 'quantize distance'."* (Callender / Huggett (2001 a) 5)

The non-renormalizability of *Covariant Quantization*, actually, is not much of a surprise – not only because it tries to 'quantize distance'. The background-independence of General Relativity, together with its identification of the gravitational with the metrical field, make a background-dependent approach to a theory of Quantum Gravity highly questionable. *Covariant Quantization* tries to quantize a background-independent theory – General Relativity – by means of a necessarily background-dependent method. Its non-renormalizability is a consequence of exactly this problem. The self-interaction of the graviton, leading to the non-renormalizability of *Covariant Quantization*, is a quantum field theoretical expression of the background-independence of General Relativity. Gravitons are theoretical artifacts, resulting from a conceptually inadequate methodology.

---

[11] All other interactions couple only to their 'charges', not to energy.

> *"The failure of the perturbative approach to quantum gravity in terms of linear fluctuations around a fixed background metric implies that the fundamental dynamical degrees of freedom of quantum gravity at the Planck scale are definitively not gravitons. At this stage, we do not yet know what they are."* (Loll (2007) 2)

So, *Covariant Quantization* shows explicitly that it is not possible to get over the mutual incompatibility of General Relativity and Quantum Mechanics / Quantum Field Theory by simply amalgamating gravity and the quantum. The conceptual foundations of both are obviously much too different. A Quantum Field Theory of gravity does not exist, because it is not possible to quantize a background-independent theory of spacetime by means of a background-dependent approach, describing a dynamics on (an already fixed) spacetime. It is not possible to describe the quantum dynamics of spacetime on spacetime.

*

Nonetheless, this is exactly what *String Theory*[12] tries to do – although in more sophisticated way than *Covariant Quantization*. *String Theory* seems to evade – obviously with more success – the problem of the non-renormalizability of the *Covariant Quantization* scheme by means of a unification of all interactions. Instead of simply describing the dynamics of gravitons on a fixed spacetime, it describes – simply – the dynamics of one-dimensionally extended strings on a fixed spacetime. So, it does not start from a direct quantization of General Relativity, but from a quantization of the classical dynamics of a relativistic string. Gravitons turn out to be quantum states of this string. But, also in *String Theory*, these graviton states move on a fixed classical spacetime. All known formulations of *String Theory* are background-dependent; although they seem to evade the non-renormalizability problem of *Covariant Quantization*, they lead to various severe and – after more than three decades of development – still unsolved problems, not to be discussed in the present context.[13]

## 5. The Spacetime Picture of *Loop Quantum Gravity*

*Loop Quantum Gravity*[14] is a much more sophisticated attempt at a direct quantization of General Relativity than the perturbative *Covariant Quantization* approach. As a *Canonical Quantization* approach, starting from the Hamiltonian formulation of General Relativity, it is intrinsically non-perturbative. And, in particular, it is background-independent. In contrast to the old geometrodynamical[15] *Canonical Quantization* approach, which started from a quantization of a Hamiltonian formulation of General Relativity with the metric and the curvature of spacetime as basic variables, *Loop Quantum Gravity* starts from a Hamiltonian formulation of General Relativity based on the so-called Ashtekar variables[16] (a spatial SU(2) connection variable and an orthonormal triad).

The Hamiltonian formulation of General Relativity results from a splitting of spacetime into spatial hypersurfaces and a time parameter. In the case of the Ashtekar variables, it is a three-dimensional connection and a time parameter. The latter is necessary for the definition of the canonical momentum as well as for the canonical quantization procedure. The (active[17]) diffeomorphism invariance of General Relativity – the formal expression of the general covariance[18] of the classical theory, interpreted in *Loop Quantum Gravity* as a gauge invariance[19] (that has to be taken into account in the transition to the quantum theory) – translates in the Hamiltonian approach to the *constraints*.[20]

---

[12] Cf. Polchinski (2000, 2000a), Kaku (1999), Green / Schwarz / Witten (1987).

[13] For a further discussion of *String Theory* and its problems, see Hedrich (2002, 2006, 2007, 2007a)

[14] Cf. Ashtekar (2007, 2007a), Ashtekar / Lewandowski (2004), Ashtekar et al. (1992), Rovelli (1991b, 1997, 1998, 2003, 2004), Smolin (1991, 2000), Thiemann (2001, 2002, 2006), Nicolai / Peeters (2006), Nicolai / Peeters / Zamaklar (2005). For a literature survey, see Hauser / Corichi (2005).

[15] Cf. DeWitt (1967), Kuchar (1986, 1993), Ehlers / Friedrich (1994).

[16] Cf. Ashtekar (1986, 1987).

These constraints are necessary, because the plain Hamiltonian theory and its basic variables do not take into account diffeomorphism invariance. The corresponding phase space contains redundant representations of physically identical spacetimes (as well as representations of physically impossible states – states that lie outside the 'constraint surface'). The identification of equivalence classes of representations of physically identical spacetimes – equivalence classes of representations that can be transformed into each other by a diffeomorphism – (as well as the identification of physically impossible states) has to be introduced additionally, by means of the constraints. Constraints are typical for the Hamiltonian formulation of dynamics with an unphysical surplus structure. Such an unphysical surplus structure is, on the other hand, typical for systems with gauge freedom. In gauge systems, it is the gauge invariance that captures unphysical redundancies in the description of a system, in the Hamiltonian formalism it is the constraints that capture them. The constraints can be understood as generators of gauge transformations. In case of General Relativity, the corresponding gauge invariance is diffeomorphism invariance.

Gauge transformations are unobservable, and if one wants to keep up the predictive power of the theory, then 'observables' have to be gauge-invariant. Formally, in the Hamiltonian approach, this means that all observables have (weakly, i.e. on the constraint surface) vanishing Poisson brackets with all (first class[21]) constraints. In the quantum case, this translates into: all quantum observables have to commute with all quantum constraints.

Already in the geometrodynamical version of the Hamiltonian formulation of General Relativity, after the splitting of spacetime into spatial hypersurfaces and a time parameter, there are four constraints: the *scalar* or *Hamiltonian constraint* and three *momentum* or *diffeomorphism constraints*.[22] In the Ashtekar version, because of an additional redundancy connected with the new variables, one has three additional *Gauss constraints*, which generate SU(2) gauge transformations.[23] After a further modification of the classical Hamiltonian theory – a transition from Ashtekar's connection vari-ables to loop variables (Wilson loops)[24] – *Loop Quantum Gravity* starts into the quantization procedure using the Dirac quantization method[25] for constrained Hamiltonian systems.

---

[17] Active diffeomorphisms are understood as point transformations, in contrast to passive diffeomorphisms, understood as coordinate transformations.

[18] Cf. Earman (1986, 1989, 2002, 2006, 2006a), Earman / Norton (1987), Norton (1988, 1993, 2004).

[19]    *"Because active diff[eomorphism] invariance is a gauge, the physical content of [general relativity] is expressed only by those quantities, derived from the basic dynamical variables, which are fully independent from the points of the manifold."* (Rovelli (2001) 108)

[20] Cf. Henneaux / Teitelboim (1992), Govaerts (2002), Belot / Earman (1999, 2001). The (primary) constraints are a direct consequence of the transition from the Lagrangian formalism to the Hamiltonian formalism by means of a Legendre transformation.

[21] First class constraints are constraints with vanishing Poisson brackets with all other constraints.

[22]    *"The invariance of the classical theory under coordinate transformation leads to four (local) constraints: the Hamiltonian constraint [...] and the three diffeomorphism (or momentum) constraints [...]."* (Kiefer (2005) 8) – *"[...] the momentum and Hamiltonian constraints are believed to capture the invariance of general relativity under spacelike and timelike diffeomorphisms respectively."* (Callender / Huggett (2001a) 19)

[23]    *"In the connection and loop approaches, three additional (local) constraints emerge because of the freedom to choose the local triads upon which the formulation is based."* (Kiefer (2005) 9)

> *"[...] Dirac introduced a systematic quantization program. Here, one first ignores constraints and introduces a kinematic framework consisting of an algebra **a** of quantum operators and a representation thereof on a Hilbert space $H_{kin}$. This provides the arena for defining and solving the quantum constraints. When equipped with a suitable inner product, the space of solutions defines the physical Hilbert space $H_{phy}$."* (Ashtekar (2007a) 2)

Under 'solving the constraints', one understands, in the classical case, a transition from a description based on the full (unconstrained) Hamiltonian phase space, containing descriptive redundancies, to a reduced phase space that captures only the 'true' (physical) degrees of freedom of the system. In the quantum case, this corresponds to the transition from the full (unconstrained) 'kinematical' quantum mechanical Hilbert space, containing redundancies (e.g. in form of gauge symmetries), to a reduced 'physical' Hilbert space, representing only the 'true' physical states of the system. The Dirac quantization method consists in a quantization of the full Hamiltonian phase space of the classical theory – canonical commutation relations for the quantum counterparts of the classical variables, an operator algebra and, finally, the quantum counterparts of the classical constraints are to be defined – with the intention to solve the quantum constraints afterwards, and to identify thereby the true physical states.

> *"Note that, in this approach, the commutation relations are simply postulated."* (Stachel (2006) 73)

An alternative to Dirac quantization would consist in solving the constraints first, for the classical theory, and then to quantize the reduced classical theory, which, then, has no constraints any more.

> *"To pass to the quantum theory, one can use one of the two standard approaches: i) find the reduced phase space of the theory representing 'true degrees of freedom' thereby eliminating the constraints classically and then construct a quantum version of the resulting unconstrained theory; or ii) first construct quantum kinematics for the full phase space ignoring the constraints, then find quantum operators corresponding to constraints and finally solve quantum constraints to obtain the physical states. Loop quantum gravity follows the second avenue [...]."* (Ashtekar / Lewandowski (2004) 51)

But, actually, the alternative to Dirac quantization is, unfortunately, nothing more than a chimera:

> *"A distinct quantization method is the* reduced phase space quantization, *where the physical phase space is constructed classically by solving the constraints and factoring out gauge equivalence prior to quantization. But for a theory as complicated as general relativity it seems impossible to construct the reduced phase space."* (Gaul / Rovelli (2000) 9) – *"Relatively little is presently known about the structure of the reduced phase space of general relativity."* (Belot / Earman (2001) 229)

Already at this point, one could ask: Why should it be easier to solve the constraints in the quantum case? – And indeed, solving *all* the quantum constraints and finding the physical Hilbert space, and thereby the true states of *Loop Quantum Gravity*, is anything but easy. Actually, no one knows how to do it. The quantized form of the Hamiltonian constraint, the so-called *Wheeler-DeWitt equation*, is well-known for its resistance against any attempt to solve it.

---

[24] A Wilson loop is the trace of a holonomy (an integral of a connection along a closed curve). Wilson loops are gauge invariant, holonomies are not.
[25] Cf. Henneaux / Teitelboim (1992).

However, there are already very interesting results for the kinematical Hilbert space in *Loop Quantum Gravity*. For the spatial hypersurfaces, after solving only the quantum Gauss constraints, one finds a discrete, polymer-like graph structure: according to *Loop Quantum Gravity*, the discrete quantum substructure to the (spatial part of the) spacetime continuum of General Relativity.[26] This *spin network* structure represents the discrete eigenvalues of two geometric operators one can define in *Loop Quantum Gravity*: the *area* and the *volume operator*.

> *"[...] a quantum spacetime can be decomposed in a basis of states that can be visualized as made by quanta of volume (the intersections) separated by quanta of area (the links). More precisely, we can view a spin network as sitting on the* dual *of a cellular decomposition of physical space. The nodes of the spin network sit in the center of the 3-cells, and their coloring determines the (quantized) 3-cell's volume. The links of the spin network cut the faces of the cellular decomposition, and their color j determine the (quantized) areas of these faces [...]."* (Rovelli (1998) 8)

This is a rather surprising result for the kinematical level:

> *"It is somewhat surprising that an important issue such as the fundamental discreteness of space emerges already at the kinematical level. One would have instead expected that is a result that emerges from the treatment of the Hamiltonian constraint, which encodes the 'dynamical' features of Einstein's theory. The discreteness thus seems to hold for more general theories than quantum general relativity."* (Kiefer (2004 [²2007]) 194)

Up to this point, only the Gauss constraints are solved. The spin networks, as well as the related area and volume operators, are not diffeomorphism invariant; they do not commute with the other quantum constraints.

> *"Note that the area operator is not invariant under three-dimensional diffeomorphisms. [...] It does also not commute with the Hamiltonian constraint. An area operator that* is *invariant should be defined intrinsically with respect to curvature invariants or matter fields. A concrete realization of such an operator remains elusive."* (Kiefer (2005) 11)

The next step consists in solving the (spatial) diffeomorphism (or momentum) constraints. This is realized in a transition from the spin networks to the diffeomorphism invariant *S-knots*: equivalence classes of spin networks with regard to spatial diffeomorphisms.

---

[26] It has to be emphasized that the discreteness of the spin network of *Loop Quantum Gravity* is a result of the direct non-perturbative quantization of General Relativity, not a feature the theory started with.

> *"Space itself turns out to have a discrete and combinatorial character. Notice that this is not imposed on the theory, or assumed. It is the result of a completely conventional quantum mechanical calculation of the spectrum of the physical quantities that describe the geometry of space."* (Rovelli (2004) 14)

But, the discreteness of the spin networks is not that of a regular cellular arrangement or grid (like e.g. in cellular automata), but a discreteness that requires the continuum of real numbers (like Quantum Mechanics) for its definition.

> *"This discreteness of the geometry, implied by the conjunction of [general relativity] and [quantum mechanics], is very different from the naive idea that the world is made by discrete bits of something. It is like the discreteness of the quanta of the excitations of a harmonic oscillator. A generic state of spacetime will be a continuous quantum superposition of states whose geometry has discrete features, not a collection of elementary discrete objects."* (Rovelli (2001) 110)

The discreteness of spin networks presupposes the spacetime manifold of General Relativity, although *Loop Quantum Gravity* tries to discuss away the manifold after quantization, as we will see below.

> *"Let us emphasize again that the 'discreteness' of the spin networks does not correspond to a naive discretisation of space. Rather, the underlying continuum, on which the spin networks 'float', the spatial manifold Σ, is still present."* (Nicolai / Peeters / Zamaklar (2005) 18)

*"Within the framework of loop quantum gravity, regarding s-knot states, rather than spin-network states, as the genuine physical states is not an optional move that one might be persuaded to take in response to some analogue of the hole argument. A quantum theory which countenances spin-network states as physical states is simply not a quantum version of general relativity."* (Pooley (2006) 378)

S-knots are abstract topological objects – excitation states of the gravitational field – that do not live on a background space, but rather represent space itself. Although the spacetime manifold is required to derive the S-knots, they are, according to *Loop Quantum Gravity*, the entities defining space. Every localization is a localization with regard to the S-knots. According to *Loop Quantum Gravity*, space is a completely relational construct defined by the S-knots.

*"The spin network represent relational quantum states: they are not located in a space. Localization must be defined in relation to them."* (Rovelli (2001) 110) – *"[...] in quantum gravity the notion of spacetime disappears in the same manner in which the notion of trajectory disappears in the quantum theory of a particle."* (Rovelli (2004) 21)

But S-knots represent only quantum space, not spacetime. They are not invariant with regard to temporal diffeomorphisms. They are not yet the states of the true, physical Hilbert space of the theory. The necessary last step would consist in solving the quantum Hamiltonian constraint (i.e. the Wheeler-DeWitt equation). But, until now, *Loop Quantum Gravity* did not succeed with this project.[27]

*"The main open problem is the correct implementation (and solution) of the Hamiltonian constraint."* (Kiefer (2004 [²2007]) 198) – *"[...] so far the problem of finding physical observables in quantum gravity is still very little explored territory [...]."* (Gaul / Rovelli (2000) 47)

Not even the definition of the quantum Hamiltonian constraint is unambiguous.

*"[...] there is still a large number of poorly controlled ambiguities in the definition of the Hamiltonian constraint."* (Ashtekar (2007a) 12)

And there are further serious problems in *Loop Quantum Gravity*: One of these, and probably the most severe, is that no low-energy approximation and no classical limit have been derived until now.

*"The main difficulties of loop quantum gravity lie in recovering low energy phenomenology. Quantum states corresponding to the Minkowski vacuum and its excitation have not yet been constructed, and particle scattering amplitudes have not been computed."* (Rovelli (2007) 1301)

Especially, it was not possible to derive the Einstein field equations (or anything similar to them) as a classical limit.

*"[...] there remain however, hard issues concerning whether and how classical general relativity dominates a suitably defined low energy limit. The fact that the theory is well defined and finite does not, so far as we know, guarantee that the low energy limit is acceptable."* (Smolin (2003) 27)

---

[27] Some insiders do not even expect (any more) a complete solution to this problem:
*"The final step [...] remains to be done: the physical states of the theory should lie in the kernel of the quantum Hamiltonian constraint operator. Of course we do not expect to find a complete solution of the Hamiltonian constraint, which would correspond to a complete solution of the theory."* (Gaul / Rovelli (2000) 39)
Only treatments with many simplifications exist.
*"A more complete treatment would include exponentially increasing efforts [...]."* (Gaul / Rovelli (2000) 41)

Here, one should remember that it is not a necessary requirement for a theory of Quantum Gravity to quantize General Relativity in a conceptually coherent way (although this seems to be a natural strategy). Rather, the basic and indispensable requirement for such a theory is that it is able to reproduce the phenomenology of gravity: the classical, low-energy case. Should it not be possible to do this, this would be the end of *Loop Quantum Gravity*.

> *"Loop quantum gravity [...] will fail if it turns out that the low energy limit of quantum general relativity coupled to matter is not classical general relativity coupled to quantum matter fields."* (Smolin (2003) 32)

Although it is still unclear at the moment, if *Loop Quantum Gravity* will finally succeed in the reproduction of the phenomenology of gravity, it is already totally clear that it has radical implications in comparison to the well-established theories of physics.

> *"[...] the theory gives up unitarity, time evolution, Poincaré invariance at the fundamental level, and the very notion that physical objects are localized in space and evolve in time."* (Rovelli (2007) 1302)

Probably the most radical of its consequences is the *problem of time*. It is already present in General Relativity, but it has more severe implications in *Loop Quantum Gravity*. – In General Relativity, coordinate time is not diffeomorphism invariant. According to the gauge-theoretical interpretation[28] of (the Hamiltonian formulation of) General Relativity, it is a gauge variable. The Hamiltonian constraint, capturing the transition from one spatial hypersurface to another, and therefore the dynamics of the system, can be understood as a gauge transformation.

> *"This means that each dynamical trajectory lies in a single gauge orbit: as the gravitational field evolves, it always stays in the same gauge orbit."* (Belot / Earman (2001) 225)

Essentially, this is nothing more than a circumscription of the fact that, as a result of the diffeomorphism invariance of the theory, dynamical transitions, generated by the Hamiltonian constraint, do not lead to any observable consequences. – So, because it is not diffeomorphism invariant, coordinate time is unobservable in General Relativity. And clock time, as an observable physical quantity, is a non-trivial function of the gravitational field, leading to such effects as the clock paradox. There does not exist any external, observable time parameter in General Relativity. If only observable quantities are taken into account as relevant physical quantities, then General Relativity is a theory without time and without temporal evolution. This is finally a consequence of general covariance, captured in the diffeomorphism invariance of the theory.

> *"[General Relativity] does not describe evolution with respect to an external time, but only relative evolution of physical variables with respect to each other. In other words, temporal localization is relational like spatial localization. This is reflected in the fact that the theory has no hamiltonian (unless particular structures are added), but only a 'hamiltonian' constraint. "* (Rovelli (1998) 20)

But, in the classical case, the practical consequences of the *problem of time* are limited:

> *"Such a weakening of the notion of time in classical [general relativity] is rarely emphasized, because, after all, in classical physics we may disregard the full dynamical structure of the dynamical theory and consider only a single solution of its equations of motion. [...] a single solution of the [general*

---

[28] Cf. Earman (2006, 2006a), Belot / Earman (1999, 2001).

*relativistic] equations of motion determines a spacetime, where a notion of proper time is associated to each timelike worldline."* (Rovelli (2007) 1318)

This is different for the quantum case.

*"In the quantum context, on the other hand, there is no single spacetime, as there is no trajectory for a quantum particle, and the very concept of time becomes fuzzy."* (Rovelli (2007) 1318)

After the canonical quantization of General Relativity, there are no fundamental equations that describe a temporal evolution of the system. This is because of the fact that the temporal evolution of the system is coded into the Hamiltonian constraint, which generates a gauge transformation. The corresponding gauge symmetry reflects nothing more than a descriptive redundancy of the theory, something with no observable physical counterpart. So, the quantized Hamiltonian constraint makes *Loop Quantum Gravity* a theory without time.

*"Since the quantum Hamiltonian is zero, there is no evolution in time of the quantum states. This is the core of the* problem of time*: there appears to be no time or change in quantum gravity."* (Belot / Earman (1999) 176)

All observables of the theory are timeless, because all corresponding quantum operators have to commute with the quantum Hamiltonian constraint, into which the temporal evolution of the system is coded.

*"The definition of 'observable' in the context of constrained systems is given as a variable that (weakly) commutes with all the first class constraints. However, since one of these is the generator of time evolution (the Hamiltonian constraint), the observables must be constants of motion."* (Rickles (2005) 12)

And nothing can change this fact, if one is decided only to accept observable quantities as physically relevant, in other words, if one is decided only to accept gauge-invariant operators: quantum observables which commute with all quantum constraints. – But, the *problem of time* is in direct conflict with the world of changes we perceive obviously. There are different attempts at a solution of the *problem of time* in *Loop Quantum Gravity* (as well as in the context of the geometrodynamical version of the *Canonical Quantization* approach) – all are radical and some are rather obscure – which we will not discuss here.[29]

## 6. Gravity as an Intrinsically Classical Phenomenon ?

Why did not even one of the attempts to quantize gravity lead to a successful construction of a theory of Quantum Gravity, although the first of these attempts go back to the thirties and the forties of the 20th century? Why did all attempts to quantize gravity lead to serious conceptual problems? Why do they have problems to reconstruct General Relativity as a classical limit? Why does gravity pose such problems to its quantization? In other words: What is so special about gravity? – There could be a simple explanation for the problems with the attempts to quantize gravity:

---

[29] The main approaches are Barbour's *timeless universe* (Cf. Barbour (1994, 1999), see also Butterfield (2002)), Kuchar's attempt to restore time by treating the Hamiltonian constraint differently from the other constraints (Cf. Kuchar (1991, 1992)), and Rovelli's *relational time* approach (Cf. Rovelli (1991, 1991a, 2001, 2001a, 2002, 2004, 2007)).

*"[...] gravity could all in all be an intrinsically classic / large scale phenomenon [...]."* (Girelli / Liberati / Sindoni (2008) 1)

But, if gravity is an intrinsically classical phenomenon, what about the arguments against semi-classical theories of gravity? – Those arguments presuppose that gravity is a fundamental interaction. They lose their validity if gravity is not fundamental, if it does not even appear in a fundamental quantum description of nature. Then, on the fundamental level, there is no semi-classical hybrid dynamics that would lead to conceptual inconsistencies. So, if gravity is an intrinsically classical phenomenon, it can not be a fundamental interaction. It has to be an induced or residual effect, caused by a quantum substrate dominated by other interactions. This quantum substrate would not include gravity, but would induce gravity on a higher structural level, for small energies or for long distances.[30] So, an intrinsically classical gravity has to be an emergent phenomenon that does not exist on the fundamental level of the quantum substrate.

This emergence scenario does not only reconcile an intrinsically classical gravity with the known arguments against semi-classical theories of gravity; it negates at the same time any motivation for a quantization of gravity as a means to get over the (apparent) incompatibilities between General Relativity and Quantum Mechanics. If gravity is not a fundamental interaction, it has not to be quantized to make it compatible with Quantum Mechanics. Resulting as a classical phenomenon from a quantum substrate, it would already by compatible with Quantum Mechanics. And it would not only be unnecessary to quantize gravity – rather it would be completely erroneous. A quantization of gravity would be a quantization of collective, non-fundamental, emergent degrees of freedom. This would explain very well the failure of all attempts to quantize gravity.

Under these conditions, the strategy for the development of a theory of 'Quantum Gravity' – a theory which would dispel the apparent incompatibility between General Relativity and Quantum Mechanics – would rather consist in the search for an adequate quantum substrate, and for a theory that would explain how the dynamics of this quantum substrate leads to an emergent level with an intrinsically classical gravity, having the known phenomenology. Then, the search for a theory of 'Quantum Gravity' does not mean any more a search for a theory which tries to identify the quantum properties of gravity, but for a theory which identifies the quantum substrate from which gravity emerges as a purely classical phenomenon.

## 7.     Emergent Gravity and/or Emergent Spacetime

If gravity is an intrinsically classical, residual or induced, emergent phenomenon, without any quantum properties: What about spacetime? – If General Relativity gives an adequate description of classical gravity, the general relativistic relation between gravity and spacetime, i.e. the geometrization of gravity, should be taken seriously, at least as long as no better reasons make this questionable. General Relativity would have to be seen as a classical, low-energy, long-distance limit to a searched-for theory describing the quantum substrate, from which gravity *and* spacetime results. But, then, this substrate would neither contain gravity, nor would it presuppose spacetime, at least not the continuous, dynamical spacetime of General Relativity, into which the gravitational field is encoded as metric field. The spacetime of General Relativity – we would have to expect – would be, like gravity, an emergent phenomenon. It would not be fundamental, but the macroscopic result of the dynamics of a non-spacetime ('pregeometric'[31]) substrate.

---

[30] As a first idea with regard to this emergence of gravity, one could think possibly of an analogy to the emergence of *Van der Waals forces* from electrodynamics.
[31] 'Pregeometric' does not necessarily mean 'non-geometric', but 'pre-general-relativistic-spacetime-continuum'.

*"Within the [emergent gravity framework] the very concepts of geometry and gravitational interaction are not seen as elementary aspects of Nature but rather as collective phenomena associated to the dynamics of more fundamental objects."* (Girelli / Liberati / Sindoni (2008) 1)

There is already a convincing argument for the existence of discrete microscopic degrees of freedom below the level of a continuous spacetime. It comes from the *Bekenstein-Hawking entropy* of black holes.[32] The *Covariant* (or *Holographic*) *Entropy Bound*[33], which can be motivated within the thermodynamics of black holes, can be seen as an indication for a finite information content of any spacetime volume – a finite number of degrees of freedom within a spacetime region –, which is in direct contradiction to a continuous spacetime and to the idea of fields defined on this continuous spacetime, fields that imply an infinite number of degrees of freedom for any spacetime region. – This argument for a finite information content of any finite spacetime region can be read as an indication either for a discrete spacetime structure or for a finite pregeometric structure of micro-constituents, from which spacetime results. The first alternative, that spacetime has a discrete quantum substructure, i.e. that spacetime has quantum properties leading to a finite information content, finds one of its best realizations in the spin networks at the kinematical level of *Loop Quantum Gravity*. But, the, at best, only very limited success of all attempts to quantize gravity and spacetime makes this first alternative less probable. So, the best explanation for the finite information content can be seen in the second alternative; it would then to be read as an indication for a (with regard to its degrees of freedom) finite pregeometric microstructure from which spacetime emerges.

But from which structure do gravity and spacetime emerge? Of what entities and interactions does the substrate consist?[34] Does matter (and do other quantum fields) also emerge from the substrate? – Meanwhile, there exist a lot of different, more or less (mostly less) convincing scenarios that try to answer these questions; some are conceptually interrelated and some are completely independent. Some of these scenarios take General Relativity as an adequate description of gravity and spacetime – as an effective theory for the macroscopic, low-energy regime –, keep to the general relativistic relation between gravity and spacetime, and treat them as emerging together from a pregeometric substrate. Others take General Relativity as a theory with limited validity, even for the classical, macroscopic regime – especially with regard to its geometrization of gravity –, and describe the emergence of gravity from a substrate that already presupposes spacetime. Some are pregeometric with regard to space, but not with regard to time, which is presupposed, either as a continuous parameter, or in form of discrete time steps. Most of the scenarios presuppose the validity of Quantum Mechanics on the substrate level, but a few try also to explain the emergence of Quantum Mechanics from a (sometimes deterministic) pre-quantum substrate. – Here is a selection:

(a)     *Space(time) as an expression of a spectrum of states*
        *of pregeometric quantum systems:*

In the scenario of Kaplunovsky and Weinstein[35] (which does not even mention gravity), space and its dimensionality and topology are dynamical results of the formation of higher-level order parameters within the spectrum of states constituting the low-energy regime of relatively simple pregeometric quantum systems.[36]

---

[32] Cf. Bekenstein (1973, 1974, 1981, 2000, 2001), Wald (1994, 2001), Bousso (2002).

[33] Cf. Bekenstein (1981, 2000, 2001), Bousso (2002), Pesci (2007, 2008).

[34] Certainly, it won't be fields, because they presuppose an infinite information content as well as a continuous spacetime on which they are defined.

[35] Cf. Kaplunovsky / Weinstein (1985); see also Dreyer (2004).

[36] Fermionic degrees of freedom lead to a flat space; bosonic degrees of freedom lead to a rolled-up space.

*"[...] the space-time continuum as an illusion of low-energy dynamics."* (Kaplunovsky / Weinstein (1985) 1879) − *"[...] dimension can be thought of as an integer-valued order parameter which characterizes distinct phases of a single dynamical system."* (Kaplunovsky / Weinstein (1985) 1895)

The quantum system, originally pregeometric, has a geometric low-energy phase.[37] Additional gauge degrees of freedom can also remain for the low-energy regime:

*"[...] residual interactions among the low-energy degrees of freedom which have the structure of a gauge theory."* (Kaplunovsky / Weinstein (1985) 1896)

In this model, the distinction between 'geometric' and 'internal' degrees of freedom can be seen as a low-energy artifact that has only phenomenological relevance. Space is finally nothing more than a fanning out of a quantum mechanical state spectrum. It is the expression of a quantum system having a low-energy state spectrum that shows a phenomenology, which can be interpreted best in a geometrical way. – But, the model, based on standard Quantum Mechanics, presupposes an external time parameter, which is finally incompatible with General Relativity.

*"There seem to be quantum systems which start out with a well-defined notion of time but no notion of space, and dynamically undergo a transition to a space-time phase [...]."* (Kaplunovsky / Weinstein (1985) 1879)

However, meanwhile, first ideas are arising with regard to the question how a temporal dynamics could emerge from a timeless 'dynamics'.[38] Such a timeless 'dynamics' would probably even have empirically testable consequences:

*"[...] the invariance under Lorentz transformations is only an approximate property of the field equations [...]. [...] our theory will show aether effects beyond second order."* (Girelli / Liberati / Sindoni (2008) 4)

(b)    *Spacetime and gravity as emergent thermodynamic or statistical phenomena*:

Jacobson[39] has shown that the Einstein field equations can be derived from a generalization of the proportionality between entropy and horizon area for black holes (Bekenstein-Hawking entropy) under consideration of the thermodynamical relation between heat, temperature and entropy.

*"It is difficult to resist concluding [...] that the horizon entropy density proportional to area is a more primitive concept than the classical Einstein equation, which appears as a thermodynamic consequence of the interplay of entropy and causality."* (Jacobson / Parentani (2003) 337)

---

*"[...] if the system was dominated by bosonic rather than fermionic fields then space-time would curl up instead of flattening."* (Kaplunovsky / Weinstein (1985) 1896)

[37] Phase transitions between spacetimes of different dimensionality are to be expected under specific dynamical conditions.

*"The existence of these phases implies the possibility that finite-temperature effects can cause dimension-changing phase transitions."* (Kaplunovsky / Weinstein (1985) 1893)

[38] Cf. Girelli / Liberati / Sindoni (2008).

[39] Cf. Jacobson (1995, 1999), Eling / Guedens / Jacobson (2006), Jacobson / Parentani (2003). See also Padmanabhan (2002, 2004, 2007).

For the derivation of the Einstein equations, temperature has to be interpreted as Unruh temperature of an accelerated observer within a local Rindler horizon. Heat is to be interpreted as energy flow through a causal horizon in the past, leading to a curvature of spacetime, corresponding to a gravitational field. But, General Relativity, derived from thermodynamics, is probably only valid under equilibrium condition:

> *"[...] one might expect that sufficiently high frequency or large amplitude disturbances of the gravitational field would no longer be described by the Einstein equation, not because some quantum operator nature of the metric would become relevant, but because the local equilibrium conditions would fail. It is my hope that [...] we shall eventually reach an understanding of the nature of 'non-equilibrium spacetime'."* (Jacobson (1995) 7)

The fundamental dynamics behind the causal horizon, from which the energy flow results, is unobservable in principle, and therefore unknown. Knowledge about this fundamental dynamics is not necessary for the derivation of the Einstein equations. They are generic. Nothing about the fundamental dynamics can be inferred from them. Indications independent from General Relativity are necessary.

(c)  *Gravity and/or spacetime as emergent hydrodynamic or condensed matter phenomena*:

Hydrodynamic and condensed matter models for emergent gravity go back to – and are partially inspired by – Sakharov's *Induced Gravity* scenario[40] of the sixties, which takes gravity as a residual effect of electromagnetism, induced by quantum fluctuations. According to this model, gravity results from Quantum Electrodynamics in the same way as hydrodynamics results from molecular physics; the Einstein-Hilbert action of General Relativity would be an approximate implication of the effective action of a Quantum Field Theory.

In Hu's model,[41] on the other hand, spacetime is taken to be a collective quantum state of many microconstituents, forming a macroscopic quantum coherence, comparable to a Bose-Einstein condensate.[42]

> *"In our view general relativity is the hydrodynamical (the low energy, long wavelength) regime of a more fundamental microscopic theory of spacetime, and the metric and the connection forms are the*

---

[40] Cf. Sakharov (2000). See also Visser (2002), Barcelo / Liberati / Visser (2005), Weinfurtner (2007).

[41] Cf. Hu (2005). See also Hu / Verdaguer (2003, 2004, 2008), Oriti (2006).

[42] Hu emphasizes that taking hydrodynamic and condensed matter models for emergent gravity serious would mean a major change of strategy for Quantum Gravity:

> *"This view marks a big divide on the meaning and practice of quantum gravity. In the traditional view, quantum gravity means quantizing general relativity, and in practice, most programs under this banner focus on quantizing the metric or the connection functions. Even though the stated goals of finding a microstructure of spacetime is the same, the real meaning and actual practice between these two views are fundamentally different. If we view [general relativity] as hydrodynamics and the metric or connection forms as hydrodynamic variables, quantizing them will only give us a theory for the quantized modes of collective excitations, such as phonons in a crystal, but not a theory of atoms or [Quantum Electrodynamics]. [...] we find it more useful to find the micro-variables than to quantize macroscopic variables."* (Hu (2005) 2)

In the context of such a strategy change, the cosmological constant problem – a direct consequence of the idea that spacetime has quantum properties, which dominate the energy of the vacuum – could be solvable in a natural way:

> *"One obvious phenomenon staring at our face is the vacuum energy of the spacetime condensate, because if spacetime is a quantum entity, vacuum energy density exists unabated for our present day late universe, whereas its origin is somewhat mysterious for a classical spacetime in the conventional view."* (Hu (2005) 4) *collective variables derived from them. At shorter wavelength or higher energies, these collective variables will lose their meaning, much as phonon modes cease to exist at the atomic scale."* (Hu (2005) 2)

According to Volovik's much more advanced scenario[43], gravity and spacetime could be emergent phenomena resulting from excitation states of a fermionic system with Fermi point (i.e. a topological defect in momentum space). These systems belong to a universality class showing low-energy behavior, which reproduces the Standard Model of Quantum Field Theory as well as the phenomenology of gravitation. They contain chiral fermions as low-energy quasi-particles as well as collective bosonic excitation states of the Fermi quantum liquid, and they lead to effective gravitational and gauge fields with their corresponding symmetries.

> *"The quasiparticles and collective bosons perceive the homogeneous ground state of condensed matter as an empty space – a vacuum – since they do not scatter on atoms comprising this vacuum state: quasiparticles move in a quantum liquid or in a crystal without friction just as particles move in empty space. The inhomogeneous deformation of this analog of the quantum vacuum is seen by the quasi-particles as the metric field of space in which they live. It is an analog of the gravitational field."* (Volovik (2003) 3)

Unfortunately, the identification of the concrete substrate – one of the main objectives of a theory of Quantum Gravity – is difficult within Volovik's condensed matter approach to emergent gravity. The best one can achieve is the identification of a universality class, from which the known low-energy phenomenology can be reproduced. But such a universality class contains, in general, completely different dynamical systems, which all lead to the same low-energy phenomenology.[44]

In the Fermi-point model, the emergent, effective spacetime is naturally four-dimensional and can have curvature, black holes and event horizons.[45] But, the equivalence principle and the general covariance of General Relativity are only approximately valid. Volovik's idea is that this is not necessarily a weakness of the theory. Possibly General Relativity contains theoretical artifacts without counterparts in reality. Its diffeomorphism invariance, representing the general covariance of the theory, could be such an artifact, ultimately going beyond the empirically tested phenomenology of gravity.

> *"The effective gravity may essentially differ from the fundamental gravity even in principle. Since in the effective gravity the general covariance is lost at high energy, the metrics which for the low-energy observers look as equivalent, since they can be transformed to each other by coordinate transformations, are not equivalent physically. As a result, in emergent gravity some metrics, which are natural in general relativity, are simply forbidden. [...] Some coordinate transformations in [general relativity] are not allowed in emergent gravity; [...] The non-equivalence of different metrics is especially important in the presence of the event horizon."* (Volovik (2007) 6)

<p style="text-align:center">*</p>

Actually, it is unclear at the moment, to what extent the hydrodynamic and condensed matter models of an emergent gravity are in conflict with basic conceptual implications of General Rela-tivity, e.g. what kind of background they need, and if they necessarily need an external time parameter or a quasi-local change rate. Could the background-independence of General Relativity, finally, be just a theoretical artifact, as some of the emergent gravity scenarios suggest? Could, finally, gravity be emergent, but spacetime fundamental? – For an emergent gravity model, a possible background-dependence would at least be less problematic than for an approach starting from a direct quantization of General Relativity (as long as there is no conflict with known phenomenology). In the direct quantization approach instead, background-dependence would be a conceptual contradiction: a background-dependent quantization of a background-independent theory. For emergent gravity there could still be reasons to take the background-independence of General Relativity as a theoretical artifact. But it would have to be very good reasons.

---

[43] Cf. Volovik (2000, 2001, 2003, 2006, 2007, 2008). See also Finkelstein (1996), Zhang (2002), Tahim et al. (2007), Padmanabhan (2004), Eling (2008).

[44] In this sense, it is comparable to the landscape of String Theory. Cf. Hedrich (2006) and the references therein.

[45] Volovik's model leads – like Hu's – to a natural explanation for a small cosmological constant, as well as for the flatness of the universe.

(d)     *Spacetime as a phenomenological result of a computational process*:

As the model to be presented in the next chapter will show: If spacetime should be an emergent information-theoretical phenomenon, some of the problematic implications of the hydrodynamic and condensed matter models, e.g. their possible inability to achieve background independence, can be avoided. – But many alternative scenarios of an information-theoretical emergence of gravity and/or spacetime with different substrate constructions (and sometimes with their own specific problems) exist. Most[46] presuppose quantum principles, but some[47] start from a non-quantum substrate and try not only to elucidate the emergence of gravity and spacetime, but also to reconstruct Quantum Mechanics as an emergent phenomenon.

Already in Wheeler's agenda for a future physical theory, destined to overcome the mutual incompatibility between General Relativity and Quantum Mechanics, one can find the following recommendation:

> "*Translate the quantum versions of string theory and of Einstein's geometrodynamics from the language of continuum to the language of bits.*" (Wheeler (1989) 362)

This is the central idea of Wheeler's *It from bit* concept[48]: going beyond spacetime to a truly pre-geometric substrate, constituted by pure information. Lloyd[49] modifies this in his *Computational Universe* approach to an *It from qubit*: Spacetime is here to be reconstructed as an emergent result of a completely background-independent quantum computation[50] – a background-independent quantum computer.[51]

> "*Because distances are derived from dynamics, without reference to an underlying spacetime manifold, the resulting theory is intrinsically covariant and background independent.*" (Lloyd (2005) 2)

---

[46] Cf. Lloyd (1999, 2005, 2007), Hsu (2007), Livine / Terno (2007), Zizzi (2001, 2004, 2005), Hardy (2007).

[47] Cf. Cahill (2002, 2005), Cahill / Klinger (1996, 1997, 1998, 2005), Requardt (1996, 1996a, 2000).

[48] Cf. Wheeler (1979, 1983, 1989).

[49] Cf. Lloyd (2005). See also Lloyd (1999, 2007).

[50] Quantum computations are superpositions of computational histories. The transition from these superpositions to a classical macroscopic spacetime consists in their decoherence.
"*The visible universe that we see around us presumably corresponds to one such decoherent history.*" (Lloyd (2005) 21)

[51] The metric of spacetime is, according to Lloyd, a direct result of the fundamental quantum computations.
"*The information that moves through the computation effectively 'measures' distances in spacetime in the same way that the signals passed between members of a set of GPS satellites measure spacetime.*" (Lloyd (2005) 7) –
"*[...] distances are quantities that are derived from the underlying dynamics of quantum systems.*" (Lloyd (2005) 2)

And because of the background-independence of the substrate, emergent spacetime fulfills – as Lloyd suggests – necessarily the Einstein field equations in their discrete form as Einstein-Regge equations.[52]

> *"Since general covariance [...] implies Einstein's equations, the geometry induced by the computational universe obeys Einstein's equations (in their discrete, Regge calculus form)."* (Lloyd (2005) 7)

But, as in almost all emergent gravity / emergent spacetime scenarios, the concrete substrate dynamics, finally, remains obscure. For the *Computational Universe* approach this means: It is unknown, on which concrete computation our universe with its specific spacetime chronogeometry is based.[53]

> *"Every quantum computation corresponds to a family of metrics, each of which obeys the Einstein-Regge equations. But which computation corresponds to the universe we see around us? What is the 'mother' computation? We do not know."* (Lloyd (2005) 23)

<p style="text-align:center">*</p>

But, independently of the problem of the identification of the substrate, the question remains: How can spacetime emerge from something so completely different from spacetime: quantum information, information flow, or basic causal relations? How can the chronogeometry of spacetime emerge from something completely pregeometric? This is probably one of the most fundamental questions to be posed with regard to the information-theoretical scenarios for the emergence of spacetime. The question results from the obvious conflict of these scenarios with out intuitions about spacetime. – A possible reconciliation with our intuitions comes from the *Holographic Screens* idea:[54]

Take an acyclic network (a graph) of directed relations ('lines') between elementary quantum systems ('vertices') without any (continuous, metrical) spacetime background. The directed relations are instantiated by flows of quantum information between the elementary quantum systems (and can be interpreted as causal relations). Dynamical changes occur locally in discrete steps. There are no continuous spacetime degrees of freedom on the fundamental level. – Then define screens that separate adjacent parts of the network, cutting through some of the lines of the network. For each screen a specific quantum information flow capacity can be found.

The crucial idea of the *Holographic Screens* concept starts from an inversion of the central implications of the Bekenstein-Hawking entropy:[55] According to Bekenstein, the entropy of a black hole is proportional to the area of its event horizon. And, according to the Holographic (or Covariant) En-

---

[52] Especially, the model makes understandable the back-reaction of the emergent spacetime metric on (computational) matter:

> *"The computational universe model is intrinsically a theory of quantum matter* coupled *to gravity, and not a theory of either quantum matter or quantum gravity on its own."* (Lloyd (2005) 13)

[53] It might even be that it is a superposition of all possible quantum computations from which our universe (or a multiverse to which it belongs) results.

> *"An appealing choice of quantum computation is one which consists of a coherent superposition of all possible quantum computations [...]."* (Lloyd (2005) 23)

Lloyd and some other investigators (Cf. e.g. Wolfram (2002); see also Poundstone (1985)) take quantum cellular automata to be the best candidates for a concretization of the quantum-computational scenario. – The question, how the possibly necessary assumption of time steps for the quantum computational substrate can be made compatible with background-independence, will be discussed in the next chapter.

[54] Cf. Markopoulou / Smolin (1999).

[55] Cf. Bekenstein (1973, 1974, 1981, 2000, 2001), Wald (1994, 2001), Bousso (2002).

tropy Bound,[56] this Bekenstein-Hawking entropy defines the maximum information content of the corresponding volume. So, the maximum information (corresponding to the number of independent degrees of freedom) contained within a spacetime volume is finite and proportional to the area of the surface of the spacetime volume.

The inversion of this Holographic Bound – the core of the *Holographic Screens* concept – consists now in the idea that the amount of quantum information that can flow through a screen – the quantum information flow capacity of the screen – defines the *area* of the screen.

> *"This leads us to suggest that the Bekenstein bound may be inverted and* area be defined to be a measure of the capacity of a screen for the transmission of quantum information.*"* (Markopoulou / Smolin (1999) 3)

And then, after having defined area as information flow capacity, a spacetime geometry can be established by means of a (secondary) network of 'holographic screens', to be defined on the (primary) network of elementary quantum systems and their causal relations.[57] So, the *Holographic Screens* concept exemplifies, how Wheeler's *It from bit* – modified to an *It from qubit* – could work in principle.

# 8. The Paradigmatic Case for Emergent Gravity and Emergent Spacetime: *Pregeometric Quantum Causal Histories*

A theoretical approach that goes by the name *Pregeometric Quantum Causal Histories*[58] can be seen at the moment as the probably most clear-cut, paradigmatic case of an attempt to construct a theory of 'Quantum Gravity' that can explain, how gravity as well as spacetime – here both have no quantum properties, because they are intrinsically classical phenomena – could emerge from a 'pregeometric' quantum substrate, presupposing for the substrate only very simple basic constituents and dynamics. This approach is worth a more extended discussion, not at least, because it succeeds in avoiding the most prominent problems of the direct quantization approaches, as well as those of most other emergent gravity / emergent spacetime scenarios.

> *"It is peculiar that the approaches that advocate that gravity is only an effective theory (string theory, condensed matter) are based explicitly on a spacetime being present while approaches that are background independent consider gravity to be fundamental. / Here, we will advocate an approach orthogonal to the quantum field theory-like approaches above (we are background independent) but also orthogonal to the usual background independent approaches (there will be no fundamental degrees of freedom for the gravitational field)."* (Markopoulou (2006) 2)

*Pregeometric Quantum Causal Histories* can not only be seen as the paradigmatic case of a pregeometric theory of Quantum Gravity, but also as a synthesis or a point of convergence of many different approaches[59] to a pregeometric quantum substrate. They are, on the one hand, a conceptual

---

extension of Sorkin's *Causal Set* approach[60], enriched by the *Holographic Screens*[61] idea and elements from Lloyd's *Computational Universe* scenario[62], which itself owes a lot to Wheeler's *It from bit*[63]. On the other hand, *Pregeometric Quantum Causal Histories* can also be seen as a generalization of causal spin networks and of the *Spin Foam*[64] approach, enriched by elements from *Algebraic Quantum Field Theory*.

The approach is at the moment completely speculative, but it gives at least an idea, how gravity and spacetime could emerge from a pregeometric substrate based exclusively on quantum information and its flow. – Its basic assumptions are:

- There is no continuous spacetime on the substrate level. The fundamental level does not even contain any spacetime degrees of freedom at all.[65]
- Causal order is more fundamental than properties of spacetime, like metric or topology.
- Causal relations are to be found on the substrate level in form of elementary causal network structures.
- Only a finite amount of information can be ascribed to a finite part of the substrate network of causal relations.[66]
- Quantum Mechanics is valid on the fundamental level.

*Quantum Causal Histories* are relational networks of quantum systems with only locally defined dynamic transitions. The basic structure is a discrete, directed, locally finite, acyclic graph. To every vertex (i.e. elementary event) of the graph, a finite-dimensional Hilbert space (and a matrix algebra of operators working on this Hilbert space) is assigned.[67] So, every vertex is a quantum system. Every (directed) line of the graph stands for a causal relation: a connection between two elementary events; formally it corresponds to a quantum channel, describing the quantum evolution from one Hilbert space to another. So, the graph structure becomes a network of flows of quantum information between elementary quantum events. *Quantum Causal Histories* are information processing quantum systems; they are quantum computers.

Because there are no spacetime degrees of freedom on the fundamental level of description, *Quantum Causal Histories* are necessarily background-independent, and therefore not in direct conceptual conflict with General Relativity. But, if this approach intends to be successful as a theory of Quantum Gravity, it has to explain *geometrogenesis*; it has to explain, how spacetime emerges from a pregeometric quantum substrate. This would be the first step on the way to a reproduction of the empirically well-tested phenomenological implications of General Relativity – the most basic and indispensable requirement for any theory of Quantum Gravity: General Relativity has at least to be reproduced as an effective theory for the macro-level. A second step would possibly consist in the explicit reproduction of the Einstein field equations as a classical, macroscopic approximation.

---

[60] Cf. Bombelli / Lee / Meyer / Sorkin (1987), Sorkin (2003), Rideout / Sorkin (2000, 2001), Rideout (2002), Henson (2006), Surya (2007).
[61] Cf. Markopoulou / Smolin (1999).
[62] Cf. Lloyd (1999, 2005, 2007).
[63] Cf. Wheeler (1989).
[64] Cf. Oriti (2001, 2003), Livine / Oriti (2003), Perez (2003, 2006), Baez (1998, 2000), Markopoulou / Smolin (1997).
[65] Then, *Quantum Causal Histories* are necessarily background-independent – and this in a much more extensive sense than General Relativity, which presupposes at least a fixed topology.
[66] This assumption is motivated explicitly by the Bekenstein-Hawking entropy (and the Holographic Entropy Bound) which leads to finite information limits for finite regions, and which could be reproduced under certain conditions even by *Loop Quantum Gravity* (cf. Meissner (2004)) and by *String Theory* (cf. Das / Mathur (2001), Lemos (2005), Peet (1998, 2001), Maldacena (1996)).
[67] This is one of the most important extensions in comparison with the *Causal Set* approach.

The basic idea (with regard to the first step: geometrogenesis) is the following: Macroscopic space-time and classical gravity do not result from a coarse-graining of quantum-geometric degrees of freedom – those do not exist according to the *Quantum Causal Histories* approach –, but from the dynamics of propagating coherent excitation states of the substrate.[68]

> "*[...] instead of looking for ways to coarse-grain the quantum geometry directly, one can first look for long-range propagating degrees of freedom and reconstruct the geometry from these.*" (Markopoulou (2006) 15) – "*[...] we will take up the idea that the effective description of a background independent theory can be characterized by the dynamics of coherent excitations in the fundamental theory and implement it by importing the method of noiseless subsystems from quantum information theory.*" (Markopoulou (2006) 3)

Then, macroscopic spacetime is necessarily dynamical, because it results from a background-independent pregeometric dynamics.[69] But, the dynamics of the effective degrees of freedom on the macro-level are necessarily decoupled from the dynamics of the substrate degrees of freedom. If they would not be decoupled, there would not be any spacetime or gravity on the macro-level, because there is none on the substrate level. In the same way, causality on the macro-level, finding its expression in the macro-level interactions, is decoupled from causality on the substrate-level. And spacetime-locality on the macro-level, if it emerges from the dynamics of coherent excitation states, has nothing to do with locality on the substrate graph structure level.[70] It is the same for time: The temporal development on the macro-level, corresponding to the dynamics of the coherent excitation states, is completely decoupled from the local temporal steps on the substrate-level.

> "*[...] truly effective spacetime means effective locality and effective time direction that are not simply Planck scale quantum corrections on the classical ones.*" (Markopoulou (2006) 29)

But what are these coherent, propagating excitation states, resulting from the substrate dynamics and leading to spacetime and gravity? And how do they give rise to spacetime and gravity? – The answer given by the *Quantum Causal Histories* approach consists in a coupling of geometrogenesis to the genesis of matter. The idea is that the coherent excitation states resulting from and at the same time dynamically decoupled from the substrate dynamics are matter degrees of freedom. And they give rise to spacetime, because they behave as if they were living in a spacetime.

> "*We propose that it is properties of the interactions of these excitations that we understand as space-time.*" (Markopoulou (2006) 2) – "*[...] all we can mean by a Minkowski spacetime is that all coherent degrees of freedom and their interactions are Poincaré invariant at the relevant scale.*" (Markopoulou (2006) 18) – "*In our approach the relationship between particles and symmetry group is exactly reversed. It is the particles that determine structures like the light cone and the symmetry group. We are thus proposing not to use the Poincaré group and its representation theory in the basic setup of the theory.*" (Dreyer (2007) 10)

---

[68] Cf. Kribs / Markopoulou (2005).

[69] However, not every pregeometric substrate has necessarily a geometric phase.

[70] "*In a given graph (the fundamental theory) there will be a notion of locality: in a graph two nodes are neighbors if they are connected by a link. We call this* microlocality. *In the known background independent theories, the dynamics is generated by moves that are local in the microscopic sense. But if this is to be a good theory, there should be a notion of classical spacetime geometry that emerges from the quantum geometry. This will give rise to another notion of locality, which we may characterize as* macrolocality. *[...] they do not coincide. [...] the notion of macrolocality should be defined directly from the interactions of the noiseless subsystems that we identify with the emergent degrees of freedom [...]* It is the fundamental evolution that is non-local with respect to our spacetime.*" (Markopoulou (2006) 24f)

So, the genesis of matter, resulting from the substrate dynamics, implies at the same time geometrogenesis. Both are inseparably coupled to each other.

"*In our view, matter and geometry have a more dual role. One can not have one without the other. Both emerge from the fundamental theory simultaneously.*" (Dreyer (2007) 4)

But, ultimately, the spacetime of the *Quantum Causal Histories* approach is nothing more than an implication of the behavior of matter. Spacetime is a completely relational construct, an expression of the phenomenology of matter dynamics. – And the matter degrees of freedom give at the same time rise to gravity, because the spacetime they bring forth by means of their behavior is a curved spacetime.[71] Gravity is nothing more than an expression of this curved spacetime.[72]

The still unproved central hypothesis of the *Quantum Causal Histories* approach is that the Einstein field equations are necessarily an implication of the dynamics of the coherent excitation states and that they can finally be derived from the substrate dynamics.

"*[...] the same excitations of the underlying system (characterizing the geometrogenesis phase transition) and their interactions will be used to define* both *the geometry and the energy-momentum tensor* $T_{\mu\nu}$. *This leads to the following Conjecture on the role of General Relativity: / If the assignment of geometry and* $T_{\mu\nu}$ *from the same excitations and interactions is done consistently, the geometry and* $T_{\mu\nu}$ *will not be independent but will satisfy Einstein's equations as identities. / What is being questioned here is the separation of physical degrees of freedom into matter and gravitational ones. In theories with a fixed background, such as quantum field theory, the separation is unproblematic, since the gravitational degrees of freedom are not really free and do not interact with the matter. In the classical background independent theory, general relativity, we are left with an intricate non-linear relation between the two sets: the Einstein equations. As the practitioners of canonical quantum gravity know well, cleanly extracting dynamical gravitational degrees of freedom from the matter is fraught with difficulties. If such a clean separation could be achieved, canonical quantum gravity would have succeeded at least two decades ago. / The new direction unifies matter and gravity in the pre-geometric phase and provides a path towards* explaining *gravity rather than just quantizing it.*" (Markopoulou (2007) 19)

But, what kind of matter does emerge from the substrate of the *Quantum Causal Histories* approach? And what is it, that stabilizes the coherent excitation states corresponding to matter? – The answer to the last question is: topology. The idea is that the coherent excitation states can be identified with stable topological knot structures: braids with crossings and twists.[73] These topological structures seem to be conserved by the substrate dynamics because of topological symmetries and corresponding topological conservation principles.

"*We have shown that braidings of graph edges are unaffected by the usual evolution moves. Any physical information contained in the braids will propagate coherently [...].*" (Markopoulou (2006) 19) – "*The states are bound here, not by fields, but by quantum topology. [...] the states are bound because there are conserved quantum numbers that measure topological properties of the states.*" (Bilson-Thompson / Markopoulou / Smolin (2006) 2)

---

[71] There are already concrete indications for a curved spacetime with Lorentz signature.

[72] And gravity has, as part of macro-causality, a finite propagation speed, because the coherent excitation states of the substrate, the matter degrees of freedom, have a finite propagation speed.
"*Once the velocities of the bound objects are no longer small we have to take into account that the change of [the state of the order parameters] is not instantaneous. Gravity here has a finite propagation speed.*" (Dreyer (2007) 8)

[73] Cf. Bilson-Thompson / Markopoulou / Smolin (2006), Bilson-Thompson (2005).

And, interestingly, the basic properties of these stable topological structures can be identified with the basic properties of elementary particles.[74]

> "*It is then possible that all the quantum numbers, including the geometric labels used in loop quantum gravity, can be regarded as composites of fundamentally topological properties.*" (Bilson-Thompson / Markopoulou / Smolin (2006) 11)

All particles of the Standard Model can be identified with specific topological structures.[75] (Naturally, the spectrum of topological structures does not contain any counterpart to the graviton. According to the *Quantum Causal Histories* approach, there are no gravitons: Gravity is an intrinsically classical, emergent phenomenon; it does not have any quantum properties or quantum constituents.) – But, what is still missing, is a dynamical explanation, which elucidates more extensively the identification of the basic properties of the stable topological structures with the basic properties of elementary particles. It should, finally, be possible to derive energy conservation principles from the dynamics of the stable topological structures, which should be translation-invariant; and this should, not at least, lead to an explanation for particle masses.

> "*Ultimately such rules have to arise from the dynamics.*" (Bilson-Thompson / Markopoulou / Smolin (2006) 7)

## 9.  Emergent Spacetime and the Search for a Theory of 'Quantum Gravity'

As the emergent spacetime / emergent gravity scenarios show, it is conceptually quite possible that spacetime and/or gravity are intrinsically classical, emergent, residual or induced, macroscopic phenomena without any quantum properties. And, if gravity should indeed be an emergent, intrinsically classical phenomenon, it would be completely nonsensical to try to quantize gravity. There would be no quantum properties of gravity, no gravitons etc. Gravity would be based on a substrate without any gravitational degrees of freedom.[76] A quantization of gravity would correspond to a quantization of collective, macroscopic degrees of freedom. A quantization of General Relativity would be the quantization of an effective theory describing the dynamics of these collective degrees of freedom. It would be as useful as a quantization of the Navier-Stokes equation of hydrodynamics. The resulting 'Theory of Quantum Gravity' would be analogous to something like a Quantum Hydrodynamics: an artificial, formal quantization of a classical theory describing collective, macroscopic degrees of freedom, without any implications for, or any clarifications with regard to, an underlying quantum substrate. It would be simply the wrong degrees of freedom, which are quantized.

---

[74] E.g., the twist of a braid structure can be interpreted as electromagnetic charge.
  "*Twist is interpreted as U(1) charge, so that a ± 2π twist in a ribbon represents charge ± e/3.*" (Bilson-Thompson / Markopoulou / Smolin (2006) 4)
There are also topological counterparts to charge conjugation, to quark colors, to parity etc.
  "*[...] parity inversion [...] for a braid is equivalent to a left-right inversion, while not affecting the handedness of any twists on the strands.*" (Bilson-Thompson / Markopoulou / Smolin (2006) 6)
[75] Cf. Bilson-Thompson / Markopoulou / Smolin (2006), Bilson-Thompson (2005), Bilson-Thompson / Hackett / Kauffman / Smolin (2008).
  "*The simplest non-trivial braids can be made with three ribbons and two crossings [...]. It is remarkable that with a single condition, these map to the first generation of the standard model.*" (Bilson-Thompson / Markopoulou / Smolin (2006) 4) – "*It is natural to hypothesize then that the second generation standard model fermions come from the next most complicated states, which have three crossings. [...] it is also proposed that the gauge vector bosons of the standard model are composite, and are represented by triplets of ribbons with no crossings. Braids with three ribbons and no crossings are mapped to the bosons of the electroweak interaction. The electroweak interactions between the fermions and the photon and vector bosons are then described by cutting and joining operations on 3-ribbon braids. These preserve the relevant quantum numbers.*" (Bilson-Thompson / Markopoulou / Smolin (2006) 8f)
[76] Should spacetime – under the assumption that it is related to gravity in the general relativistic sense – be, in the same way, a collective expression or result of completely different non-spacetime degrees of freedom, there would be no quantum spacetime, no fluctuations, no uncertainties, no superpositions of spacetime etc. (And, should spacetime *not* be related to gravity in the general relativistic sense, then there would not even be any initial reason at all to assume that spacetime could or should have any quantum properties.)

A quantization of gravity is only (but not necessarily) a reasonable strategy for the construction of a theory of Quantum Gravity, if gravity is a fundamental interaction. If it is not a fundamental interaction, the adequate strategy consists in a search for the substrate dynamics from which gravity emerges. 'Quantum Gravity' would then be the name for a theory describing this substrate and explaining how gravity (and spacetime) emerge(s) from this substrate. – One of the basic requirements for a theory describing this substrate dynamics is that it is possible to derive from it the empirically well-tested phenomenology of gravity.[77] Even better would be the prediction of small deviations from General Relativity, not yet in contradiction to the empirically confirmed phenomenology, which could be tested in future experiments.

So the present situation with regard to the different attempts to construct a theory of Quantum Gravity teaches us, that we should not cling to only one strategy, especially when this strategy meets serious problems in all its different instantiations. Rather we should take into account all consistent conceptual possibilities, even the more radical or exotic ones. Quantum Gravity, if we finally should succeed in the construction of a consistent and empirically confirmable theory, could very well lead to rather unexpected implications with regard to our view of gravity and spacetime.

---

[77] This is a basic requirement for all attempts to a theory of Quantum Gravity, also for those who try to quantize gravity directly. Not even *Loop Quantum Gravity* fulfills this requirement. A direct quantization of General Relativity does not at all imply or even guarantee the possibility of a reproduction of General Relativity, or its phenomenology respectively.